\documentclass[prb,twocolumn,showpacs]{revtex4}
\usepackage{color,soul}
\usepackage{amsmath}
\usepackage{dcolumn}
\usepackage{graphicx}
\usepackage{bm}
\usepackage{amssymb}
\usepackage{subfigure}
\begin{document}

\title{A short review of the recent progresses in the study of the cuprate superconductivity}
\author{Tao Li}
\affiliation{Department of Physics, Renmin University of China, Beijing 100872, P.R.China}

\begin{abstract}
The last 15 years have witnessed important progresses in our understanding of the mechanism of superconductivity in the high-$T_{c}$ cuprates. There is now strong evidence that the strange metal behavior is induced by the quantum critical fluctuation at the pseudogap end point, where the Fermi surface changes its topology from hole-like to electron-like. However, experiments show that the quantum critical behavior in the high-$T_{c}$ cuprates is qualitatively different from that observed in the heavy Fermion systems and the iron-based superconductors, in both of which the quantum critical behavior can be attributed to the quantum phase transition toward a symmetry breaking phase. The fact that the pseudogap exists as a spectral gap without a corresponding symmetry breaking order, together with the fact that the strange metal behavior occurs as a quantum critical behavior without a corresponding symmetry breaking phase transition, exposes the central difficulty of the field: the lack of a universal low energy effective theory description of the high-$T_{c}$ phenomenology beyond the Landau paradigm. Recent experiments imply that the dualism between the local moment and the itinerant quasiparticle character of the electron in the high-$T_{c}$ cuprates may serve as an organizing principle to go beyond the Landau paradigm and may hold the key to the mystery of the pseudogap phenomena and the strange metal behavior. 
  \end{abstract}
\maketitle

The intensive study during the last three decades on the mechanism of the superconductivity in the high-$T_{c}$ cuprates leaves us more puzzles than consensuses, expect for the d-wave pairing symmetry and its close relation with the antiferromagnetic spin fluctuation in the system. This should be mainly attributed to the anomalous normal state properties of such a strongly correlated electron system, which are beyond the description of the standard Landau theory of spontaneous symmetry breaking and the Fermi liquid theory. In this short review, we summarize the recent progresses in the study of the cuprate superconductivity made possible by the tremendous efforts devoted during the last 15 years. Here we will focus on drawing the implications of these recent progresses on the construction of a coherent picture for the high-$T_{c}$ problem, rather than providing a thorough review of the experimental literatures, which has been done in many excellent review articles on related topics\cite{Taillefer1,Keimer,Hussey,Taillefer2,Strange,Singh}.

The most prominent manifestation of the anomalous normal state property of the high-$T_{c}$ cuprates is the pseudogap phenomena and the strange metal behavior. In the early days of cuprates study, the word 'pseudogap' is used loosely to refer to the loss of low energy electron spectral weight inferred from various measurements. People do not know if and how these different manifestations of the pseudogap phenomena are related with each other, since they usually start at different temperatures. Current speculation on the origin of the pseudogap phenomena falls roughly into four categories\cite{Singh,Anderson,RVB,PDW,Sakai,topological}, in which the pseudogap is understood either as (1)a paring gap induced by superconducting fluctuation or preformed Cooper pairs in the normal state, (2) a band folding gap induced by a competing order in the particle-hole channel breaking the spatial symmetry, (3)a hybrid gap induced by an order parameter breaking both the $U(1)$ charge symmetry and the spatial symmetry, such as the pair density wave order, or, (4)a spectral feature induced by the Mott physics but not related to any particular symmetry breaking order parameter. 

To decide the origin of the pseudogap phenomena, a lot of efforts have been devoted to find the possible 'order parameter' for it. In recent years, evidence for symmetry breaking order(or low energy fluctuation of such order) has indeed been found in the pseudogap regime in various channels\cite{PDW,Comin,Frano}, such as the electron pairing, change density wave, spin density wave, charge current and the electron nematicity channel. These ordering tendencies of the pseudogap phase are collectively called intertwined orders. However, no single such symmetry breaking order is thought to have the potential to claim its primary responsibility for the origin of the pseudogap phenomena. ARPES measurements find that instead of a closed Fermi surface, there exists in the pseudogap phase only open segments of Fermi arc\cite{Arc1,Arc2,Arc3}, which does not enclose any definite volume. At the same time, it is found that the spectral weight redistribution accompanying the development of the pseudogap occurs in an energy range much more extended than the size of the pseudogap. These observations imply collectively that the pseudogap can not be understood as a single particle gap in the Fermi liquid theory framework.

Empirically, the strange metal behavior appears in the phase diagram just beyond the pseudogap regime, which implies that they are closely related with each other. ARPES measurements find that a closed Fermi surface with the correct Luttinger volume is recovered in the strange metal regime\cite{Arc1,Arc2,Arc3}. However, this does not imply that the strange metal phase is less anomalous than the pseudogap phase. In fact, there is no well defined quasiparticle peak on the literal Fermi surface in the strange metal phase. At the same time, both the resistivity and the Hall number of the system are found to increase linearly with temperature in this regime. This is to be contrasted with the $T^{2}$ behavior of resistivity and the temperature independent behavior of the Hall number we expect from the Fermi liquid theory. In addition, while a linearly temperature dependent electronic entropy is observed in the strange metal phase, as one would expect in a Fermi liquid metal, the zero temperature extrapolation of such a linear dependence generates a negative intersection. Such a strange behavior, which is dubbed as "dark entropy" by some researchers, implies that the entropy loss in the pseudogap phase is not fully recovered even above the pseudogap temperature\cite{Loram,Hussey}. In the literature, the strange metal behavior is understood either as the consequence of the electron incoherence caused by spin-charge separation and gauge fluctuation in a spin liquid background\cite{RVB}, or,  as the quantum critical behavior around a possible quantum critical point in the phase digram. According to the former scenario, the strange metal behavior should be the most evident around the optimally doped regime. The problem with such a scenario is that there is no solid evidence for electron fractionalization in the high-$T_{c}$ cuprates. According to the latter scenario, the strange metal behavior should be the most evident around the quantum critical point, most likely around $x_{c}\approx 0.19$, where the superconductivity is found to be the most robust against Zinc doping\cite{Tallon}. The problem with such a scenario is that we do not know what kind of degree of freedom is really critical at so large a doping concentration as $x_{c}$.

Recently, important clues as to the origin of the pseudogap phenomena and the strange metal behavior have emerged as a result of the systematic studies over the last 15 years. These studies, which are often multiple perspective in nature, have mapped out a detailed picture how the pseudogap phenomena evolve with both the temperature and the doping and how it is related to the strange metal behavior. Here we will only mention those observations that we think are the most relevant for the construction of a coherent  picture for the high-$T_{c}$ problem. 

Firstly, while different manifestations of the pseudogap phenomena may start at different temperature, people find that the Fermi arc phenomena occur right at the temperature where the antiferromagnetic spin fluctuation spectrum opens a gap\cite{Arc1,Arc2,Arc3}. More specifically, people find that the Fermi level crossing along $(0,\pi)-(\pi,\pi)$ disappears abruptly at the temperature where the spin relaxation rate on the Copper site($1/^{63}\mathrm{T}_{1}\mathrm{T}$) reaches its maximum\cite{Zheng}. 

Secondly, extensive resonant inelastic X-ray scattering(RIXS) measurements on the high-$T_{c}$ cuprates show that local moment fluctuation survives even in the heavily doped system, which is very remote from antiferromagnetic ordering instability\cite{RIXS1,RIXS2}. More specifically, both the dispersion and the intensity of the local moment fluctuation are found to be almost doping independent, although a moderate doping dependence in the mode broadening does exist.

Thirdly, measurements under strong magnetic field(applied to expose the possible quantum critical point buried inside the superconducting dome) find that the pseudogap phenomena ends abruptly at the doping level where the Fermi surface of the system undergoes the Lifshitz transition from a hole-like topology to electron-like topology\cite{Cv,FS}. Such a special doping is located in the overdoped side of the phase diagram and is given approximately by $x_{c}\approx 0.19$(the exact value of $x_{c}$ varies between the different cuprates families\cite{FS}). Right at such a pseudogap end point, quantum critical behavior of unknown origin is found in both resistivity and specific heat measurements. More specifically, the resistivity is found to follow a perfect linear temperature dependence from zero temperature up to very high temperature, with a slope implying a scattering rate saturating the so called Planckian limit $\frac{\hbar}{\tau}\approx  k_{B}T$\cite{Planck}. At the same time, the slope of the electronic specific heat diverges around $x=x_{c}$\cite{Cv}. 

We note that while similar quantum critical behavior has also been found in the heavy Fermion systems and the iron-based superconductors\cite{Shibauchi}, what is very different here is that no symmetry breaking quantum phase transition has been found around $x=x_{c}$. In fact, as we have stated above, no single symmetry breaking order is found to have the potential to claim its primary responsibility for the origin of the pseudogap phenomena, which can not be thought as a single particle gap in the Fermi liquid theory framework. The difference between the quantum critical behavior observed around $x=x_{c}$ and that in the iron-based superconductors shows most strikingly in the doping dependence of the superfluid density $\rho_{s}$. More specifically, while $\rho_{s}$ reaches its minimum around the quantum critical point in the iron-based superconductors, which is consistent with the divergence of the effective mass of the quasiparticle there, the pseudogap end point in the high-$T_{c}$ cuprates corresponds to the doping where $\rho_{s}$ reaches its maximum\cite{Tallon}. 

Fourthly, adding to the surprises, people find that the Hall response of the system exhibits weird behavior around $x=x_{c}$\cite{Hall}. In particular, the normal state Hall number in the zero temperature limit(achieved by applying strong magnetic field) is found to undergo a dramatic jump from the $x$ doping dependence to a $1+x$ doping dependence around $x_{c}$. In addition, it is found that while the Wiedemann-Franz law relating the thermal and charge Hall response of a Fermi liquid metal is well satisfied above $x_{c}$, it is strongly violated below $x_{c}$. More specifically, with the rapid suppression of the charge Hall conductivity below $x_{c}$, the thermal Hall conductivity is found to change its sign and increase in its magnitude monotonically with the decrease of the doping concentration, until reaching its maximum in the antiferromagnetic parent compounds\cite{THE,THEP}.

While it is not clear what is the full implications of these recent experimental findings, the following points are now clear. First, the strange metal behavior should be understood as the quantum critical behavior around the psedogap end point, although we do not know what is really critical there. It is thus better to use $x_{c}$, where $\rho_{s}$ reaches its maximum, rather than $x_{opt}$, where $T_{c}$ reaches its maximum, as the separatrix of underdoped and overdoped regime of the high-$T_{c}$ phase diagram. Such a change of focus may have important impact on the study of cuprate superconductivity. Second, the electron in the high-$T_{c}$ cuprates exhibits simultaneously the local moment and the itinerant quasiparticle character in the whole phase diagram. The tradeoff between the charge Hall response and the thermal Hall response in the pseudogap phase may simply be the consequence of the interconversion between these two characters\cite{THE3}. The situation here is very different from that in the heavy Fermion systems and the iron based superconductors, in which the local moment and itinerant quasiparticle behavior of the system can be attributed to electron occupying different orbitals. This may explain why these systems exhibit different quantum critical behaviors. Third, the antiferromagnetic fluctuation of the local moment is intimately related to the origin of the pseudogap phenomena. This can be inferred either from the fact that the Fermi arc and the antiferromagnetic spin fluctuation gap appear simultaneously with the lowering of temperature\cite{Arc1,Zheng}, or the fact that the pseudogap can be observed only on hole-like Fermi surface\cite{FS}, since the hot spot for antiferromagnetic scattering exists only on hole-like Fermi surface. 

With these understandings in mind, we are left with the following two major puzzles. First, the pseudogap exists as a spectral gap without a corresponding symmetry breaking order. Second, the strange metal behavior occurs as a quantum critical behavior without a corresponding symmetry breaking phase transition. These two puzzles, just as the smile of Cheshire cat, expose the central difficulty of the field: the lack of a universal low energy effective theory description of the high-$T_{c}$ phenomenology beyond the Landau paradigm. 

According to the Landau paradigm, an interacting Fermion system should behave like an non-interacting system at low energy if the perturbative expansion in the interaction converges, while the divergence of the perturbative expansion usually implies the spontaneous breaking of a symmetry. The perturbative expansion around the symmetry breaking saddle point usually becomes well behaved again. Such a scheme ignores an important possibility in which the divergence of the perturbative expansion does not lead to any spontaneous symmetry breaking phase, but to a highly entangled quantum fluid - the non-Fermi liquid. The lack of consensus on the mechanism of the superconductivity in the high-$T_{c}$ cuprates can be largely attributed to the lack of systematic understanding on the property of a non-Fermi liquid.

Establishing a unified low energy effective theory description for the high-$T_{c}$ phenomenology beyond the Landau paradigm is not only in urgent need in the study of this particular system, but will also contribute significantly to the study of more general strongly correlated electron systems. The key to achieve this goal is to find the organizing principle why the Landau paradigm fails in the high-$T_{c}$ cuprates. We think the local-itinerant dualism of the electron in high-$T_{c}$ cuprates may just serve as such an organizing principle. In the Fermi liquid theory, collective spin fluctuation emerges at low energy only when the system is within or in proximity to a magnetic ordered phase. When the system is far away from magnetic instability, the only low energy degree of freedom that we would expect after we integrate out the Fermion degree of freedom with large momentum is the itinerant quasiparticle. Without the Fermionic degree of freedom at large momentum, the local moment simply can not be defined. Thus for a single band electron system, the local-itinerant dualism of electron in the low energy physics can happen only when we go beyond the Landau paradigm.  

At the phenomenological level, we can describe the local-itinerant dualism of electron in the high-$T_{c}$ cuprates with the so called spin-Fermion model, which treats the itinerant quasiparticle and the local moment aspect of the same electron as two independent but coupled degree of freedom. While such a phenomenological description has been proposed in the early days of the high-$T_{c}$ study\cite{MMP,Chubukov}, its use is largely limited to the perturbative regime. In recent years, people find that some well-designed modification on the model can make it solvable by sign-free quantum Monte Carlo simulation\cite{Berg}. This gives us the hope to go systematically beyond the perturbative regime in the study of this model. 

However, is such a simple model indeed sufficient to describe the complex phenomenology of the high-$T_{c}$ cuprates? In particular, what is the origin of the quantum critical behavior around $x_{c}$ and the intertwined orders inside the pseudogap phase according to such a picture\cite{QCP}? At the same time, it is important to know how such a phenomenological description can emerge from the microscopic models of the high-$T_{c}$ cuprates. In particular, it is important to know how the evolution of the itinerant quasiparticle character and the local moment character with doping and temperature would feedback on each other. It is also interesting to know what kind of consequence would be left on the electron spectrum in the process of such an emergence. In particular, it is interesting to know what is the signature of the local-itinerant dualism in the electron spectrum of the high-$T_{c}$ cuprates\cite{signature}. To establish such a connection between the microscopic and the low energy effective picture, we need systematic knowledge on the dynamics of the charge, spin and single particle degree of freedom in the intermediate to high energy range\cite{mode}.

In this short review, we have summarized the main progresses made in the last 15 years on our understanding of the mechanism of the superconductivity in the high-$T_{c}$ cuprates. There is now strong evidence that the strange metal behavior is induced by the quantum critical fluctuation at the pseudogap end point, where the Fermi surface changes its topology from hole-like to electron-like. Experiments also show that the quantum critical behavior in the high-$T_{c}$ cuprates is qualitatively different from that observed in  the heavy Fermion systems and the iron based superconductors. The challenge raised by these new experimental findings can be summarized as the following two puzzles: the pseudogap exists as a spectral gap without a corresponding symmetry breaking order and the strange metal behavior occurs as a quantum critical behavior without a corresponding symmetry breaking phase transition. Both puzzles call for the development of a unified low energy effective theory description of the high-$T_{c}$ phenomenology beyond the Landau paradigm. The new experiment findings also imply that the local-itinerant dualism of the electron in the high-$T_{c}$ cuprates may serve as an organizing principle to go beyond the Landau paradigm and that the antiferromagnetic correlation between the local moment is crucial in resolving the mystery of pseudogap phenomena and strange metal behavior.

\begin{acknowledgments}
I thank Tian-Xing Ma and Shi-Ping Feng for inviting me to write a review article on this topic. I also acknowledge the support from the grant NSFC 11674391 and the grant National Basic Research Project 2016YFA0300504.
\end{acknowledgments}


\begin{thebibliography}{}
\bibitem{THE3}We note that recent experiment reports\cite{THEP} that the thermal Hall response in the high-$T_{c}$ cuparte is isotropic in all three space dimension. This is obviously beyond the description of any two dimensioanl model of the high-$T_{c}$ cuprates. However, we note that a transverse thermal current in the presence of a longitudinal temperature gradient does not necessarily imply a thermal Hall response. It may simply be the result of the anisotropy in the thermal conductivity tensor. This point is especially relevant when the thermal Hall response is measured with the magnetic field applied parallel to the CuO$_{2}$ plane.
\bibitem{QCP}While a full answer to these questions still await future works, we note that the quantum critical behavior at $x=x_{c}$ may be explained without invoking symmetry breaking phase transition if we assume that the local moment system(which is assumed to be far from critical but is otherwise antiferromagnetically correlated) is coupled to a quasiparticle system with a divergent antiferromagentic susceptibility. The latter is true because the Van Hove singularity and the antiferromagnetic hot spot coincide on the Fermi surface when $x=x_{c}$\cite{Tao}.
\bibitem{signature}Here what concern us are those spectral anomalies that are beyond the description of the low energy effective theory.  The local-itinerant dualism of electron in the high-$T_{c}$ cuprates is the most important origin for such spectral anomalies at the intermediate to high energy range, for example, the waterfall structure and the prominent particle-hole asymmetry in the single particle spectrum,  the ubitoqus non-Drude behavior in the optical conductivity spectrum, the Raman spectrum and the charge fluctuation spectrum, and the high energy continuum in the spin fluctuation spectrum which is up to now still poorly understood as a result of the limitation in the measurement resolution.
\bibitem{mode}Many previous studies have focused on the search of a 'paring glue' from the emergent degree of freedom at low energy, such as the neutron resonance mode, but have forgotten the fact that such emergent degree of freedom appears on the background of anomalous dynamics at the intermediate to high energy range, which may have a closer relation with the mechanism of superconductivity in the high-$T_{c}$ cupartes. A more detailed analysis on this point can be found in Ref.[\onlinecite{Tao1}].
\end{thebibliography}
\end{document}